\documentclass[pre,floatfix,twocolumn,showpacs]{revtex4}
\usepackage{amsmath}
\usepackage{amsfonts}
\usepackage{epsfig}
\usepackage{longtable}

\begin{document}

\title{Collective behavior of stock price movements in an emerging market}
\author{Raj Kumar Pan}
\email{rajkp@imsc.res.in}
\author{Sitabhra Sinha}
\email{sitabhra@imsc.res.in}
\affiliation{%
The Institute of Mathematical Sciences, C.I.T. Campus, Taramani,
Chennai - 600 113 India
}%
\date{\today}

\begin{abstract}
To investigate the universality of the structure of interactions in
different markets, we analyze the cross-correlation matrix ${\mathbf C}$ of
stock price fluctuations in the National Stock Exchange (NSE) of India. We
find that this {\em emerging} market exhibits strong correlations in the
movement of stock prices compared to {\em developed} markets, such as the
New York Stock Exchange (NYSE). This is shown to be due to the dominant
influence of a common market mode on the stock prices.  By comparison,
interactions between related stocks, e.g., those belonging to the same
business sector, are much weaker. This lack of distinct sector identity in
emerging markets is explicitly shown by reconstructing  the network of
mutually interacting stocks.  Spectral analysis of ${\mathbf C}$ for NSE
reveals that, the few largest eigenvalues deviate from the bulk of the
spectrum predicted by random matrix theory, but they are far fewer in
number compared to, e.g., NYSE. We show this to be due to the relative
weakness of intra-sector interactions between stocks, compared to the market
mode, by modeling stock price dynamics with a two-factor model. Our
results suggest that the emergence of an internal structure comprising
multiple groups of strongly coupled components is a signature of market
development.
\end{abstract}

%{89.65.Gh}{Economics;econophysics, financial markets, business and management}
%{89.65.-s}{Social and economic systems}
%{05.40.Ca}{Noise}
%{05.40.Fb}{Random walks and Levy flights}
%{05.45.Tp}{Time series analysis}
%{89.75.-k}{Complex systems}

\maketitle
\section{Introduction}
\label{sec:1}
Financial markets can be considered as complex systems having many
interacting elements and exhibiting large fluctuations in their associated
observable properties, such as stock price or market
index~\cite{Mantegna99,Bouchaud03}. The state of the market is governed by
interactions among its components, which can be either traders or stocks.
In addition, market activity is also influenced significantly by the
arrival of external information. Statistical properties of stock price
fluctuations and correlations between price movements of different stocks
have been analyzed by physicists in order to understand and model financial
market dynamics~\cite{Kondor99,Chatterjee06}. The fluctuation distribution
of stock prices is found to follow a power law with exponent $\alpha \sim
3$, the so-called ``inverse cubic law''~\cite{Lux96,Plerou99}. This
property is quite robust, and has been seen in developed as well as
emerging markets~\cite{Pan07}.  On the other hand, it is not yet known
whether the cross-correlation behavior between stock price fluctuations has
a similar universal nature. Although the existence of collective modes have
been inferred from the study of market dynamics, such studies have almost
exclusively focussed on developed markets, in particular, the New York
Stock Exchange (NYSE).

To uncover the structure of interactions among the elements in a financial
market, physicists primarily focus on the spectral properties of the
correlation matrix of stock price movements. Pioneering studies
investigated whether the properties of the empirical correlation matrix
differed from those of a random matrix that would have been obtained had
the price movements been uncorrelated~\cite{Laloux99,Plerou99_prl}. Such
deviations from the predictions of random matrix theory (RMT) can provide
clues about the underlying interactions between various stocks. It was
observed that, while the bulk of the eigenvalue distribution for the
correlation matrix of NYSE and Tokyo Stock Exchange follow the spectrum
predicted by RMT~\cite{Laloux99,Plerou99_prl,Plerou02,Utsugi04}, the few
largest eigenvalues deviate significantly from this. The largest eigenvalue
has been identified as representing the influence of the entire market,
common for all stocks, whereas, the remaining large eigenvalues are associated
with the different business sectors, as indicated by the composition of
their corresponding eigenvectors~\cite{Gopikrishnan01,Plerou02}. The
interaction structure of stocks in NYSE have been reconstructed using
filtering techniques implementing matrix decomposition~\cite{Kim05} or
maximum likelihood clustering~\cite{Giada01}. Correlation matrix analysis
has applications in the area of financial risk management, as mutually
correlated price movements may indicate the presence of strong interactions
between stocks~\cite{Markowitz59}.  Such analyses have been performed using
asset trees and asset graphs to obtain the taxonomy of an optimal portfolio
of stocks~\cite{Mantegna99_epjb,Onnela02,Onnela03}.

While it is generally believed that stock prices in emerging markets tend
to be relatively more correlated than the developed ones~\cite{Morck00},
there have been very few studies of the former in terms of analysing the
spectral properties of correlation
matrices~\cite{Wilcox07,Sinha06,Kulkarni06,Cukur07}.
Most studies of correlated price movements in emerging markets have looked
at the {\em synchronicity} which measures the incidence of similar (i.e.,
up or down) price movements across stocks~\cite{Morck00,Durnev04}. Although
related to correlation the two measures are not same, as correlation also
gives the relative magnitude of similarity. In this paper, we analyze the
cross-correlations among stocks in the Indian financial market, one of the
largest emerging markets in the world. Our study spans over 1996-2006, a
period which coincides with the decade of rapid transformation of the
Indian economy after liberalization in the early 1990s.

We find that, in terms of the properties of its collective modes, the
Indian market shows significant deviations from developed markets. As the
fluctuation distribution of stocks in the Indian
market~\cite{Pan07,Pan06,Sinha06} follows the ``inverse cubic law" seen in
NYSE~\cite{Plerou99,Gopikrishnan99}, the deviations observed in the
correlation properties should be almost entirely due to differences in the
nature of interaction structure in the two markets. Our observation of a
higher degree of correlation in the Indian market compared to developed
markets is found to be the result of a dominant market mode affecting all the
stocks, which is further accentuated by the relative absence of clusters of
mutually interacting stocks. This is explicitly shown by reconstructing the
network of interactions within the market, using a filtered correlation
matrix from which the common market influence and random noise has been
removed. This procedure give a more accurate representation of the
intra-market structure than the commonly used method of constructing
minimal spanning tree from the unfiltered correlation
matrix~\cite{Mantegna99_epjb,Onnela02,Kim05}. To support the interpretation
of our empirical observations, we present a two-factor model of market
dynamics in Section~\ref{sec:4}. Multi-factor models of market behavior
have been used by other groups for explaining various spectral properties
of empirical correlation matrices~\cite{Noh00,Lillo05,Roman06}. In this
model, we assume the market to consist of several correlated groups of
stocks which are also influenced by a common external signal, i.e., market
mode. By varying the relative strength of the factor associated with the
market mode to that associated with the groups, we show that decreasing the
intra-group interactions result in spectral distribution properties similar
to that seen for the Indian market. Our results imply that one of the key
features signifying the transition of a market from emerging to developed
status is the appearance and consolidation of distinct group identities.

\section{Data Analyzed}
\label{sec:2}
The National Stock Exchange (NSE) is the largest stock market in India.
Having commenced operations from Nov 1994, it is already the world's third
largest stock exchange (after NASDAQ and NYSE) in terms of
transactions~\cite{Nse04}. It is thus an excellent source of data for
studying the correlation structure of price movements in an {\em emerging}
market.

We have considered the daily closing price data of 201 stocks (see
Table I) traded in NSE
from Jan 1996 to May 2006, which corresponds to 2607 days.  This data is
obtained from the NSE web-site~\cite{Nse} and has been manually corrected
by us for stock splitting.  The selected stocks were traded over the entire
period 1996-2006 and had the minimum number of missing data points (i.e.,
days for which no price data is available). If the price value of a stock
is missing on a particular day, a problem common to data from emerging
markets~\cite{Wilcox07}, it is assumed that no trading took place on that
day, i.e, the price remained the same as the preceding day.  For comparison
we also consider the daily closing price of 434 stocks of NYSE belonging to
the S\&P~500 index over the same period as the Indian data. However, the
total number of working days is slightly different, viz., 2622 days. This
data was obtained from the Yahoo! Finance website~\cite{Yahoo}. In all our
analysis, while comparing with the NSE data, 
%we have randomly chosen 201 stocks from the set of 434 NYSE stocks. 
we have used multiple random samples of 201 stocks each, from the set of 434
NYSE stocks.
We verified that the results
obtained were independent of the particular sample of 201 stocks chosen.

To ensure that the missing closing prices in the Indian market data do not 
result in artifacts leading to spurious divergence from the US market, we
have also performed our analysis on synthetic US market data containing 
the same number of missing data points. Multiple sets of such data were 
generated from the actual closing price time series by randomly choosing 
the required number of data points and replacing them with the same value 
as the preceding day. The resulting analysis showed no significant difference
from the results obtained with the original US data. 
%The exercise of
%generating such artificial data sets was carried out several times
%and the resulting values for the largest eigenvalue and the number of
%eigenvalues deviating from the bulk were recorded. Results of this
%comparison are mentioned in the next section.

\section{The Return Cross-Correlation Matrix}
\label{sec:3}
\begin{figure}
\includegraphics[width=0.9\linewidth]{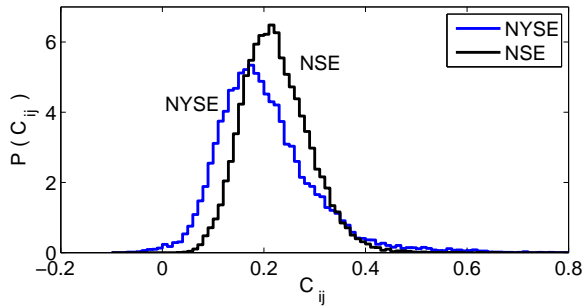}
\caption{The probability density function of the elements of the
correlation matrix ${\mathbf C}$ for 201 stocks in the NSE of India
and NYSE for the period Jan 1996-May 2006. The mean value of elements of
$\mathbf C$ for NSE and NYSE, $\langle C_{ij}\rangle$, are 0.22 and 0.20
respectively.}
\label{fig:correlation}   
\end{figure}

To observe correlation between the price movements of different stocks, we
first measure the price fluctuations such that the result is independent of
the scale of measurement. If $P_{i}(t)$ is price of the stock $i=1,\dots,N$
at time $t$, then the (logarithmic) price return of the $i$th stock over a
time interval $\Delta t$ is defined as
\begin{equation}
R_{i}(t,\Delta t) \equiv \ln {P_{i}(t+\Delta t)}- \ln {P_{i}(t)}. 
\label{eq:return}
\end{equation}
As different stocks have varying levels of volatility (measured by the
standard deviation of its returns) we define the normalized return,
\begin{equation}
r_{i}(t,\Delta t) \equiv \frac{R_{i}-\langle R_{i} \rangle}{\sigma_{i}},
\label{eq:nor_return}
\end{equation}
where 
$\sigma_{i}\equiv\sqrt{\langle R_{i}^{2}\rangle-\langle R_{i}\rangle^{2}}$, 
is the standard deviation of $R_{i}$ and $\langle \ldots \rangle$
represents time average over the period of observation. We then compute
the equal time cross-correlation matrix ${\mathbf C}$, whose element 
\begin{equation}
C_{ij} \equiv \langle r_{i} r_{j} \rangle,
\label{eq:correlation}
\end{equation}
represents the correlation between returns for stocks $i$ and $j$. By
construction, $\mathbf{C}$ is symmetric with $C_{ii}=1$ and $C_{ij}$ has a
value in the domain $[-1,1]$. Fig.~\ref{fig:correlation} shows that, the
correlation among stocks in NSE is larger on the average compared to that
among the stocks in NYSE. This supports the general belief that developing
markets tend to be more correlated than developed ones. To understand the
reason behind this excess correlation, we perform an eigenvalue analysis of
the correlation matrix. 

\subsection{Eigenvalue spectrum of correlation matrix}
\begin{figure}
\centering
\includegraphics[width=0.85\linewidth]{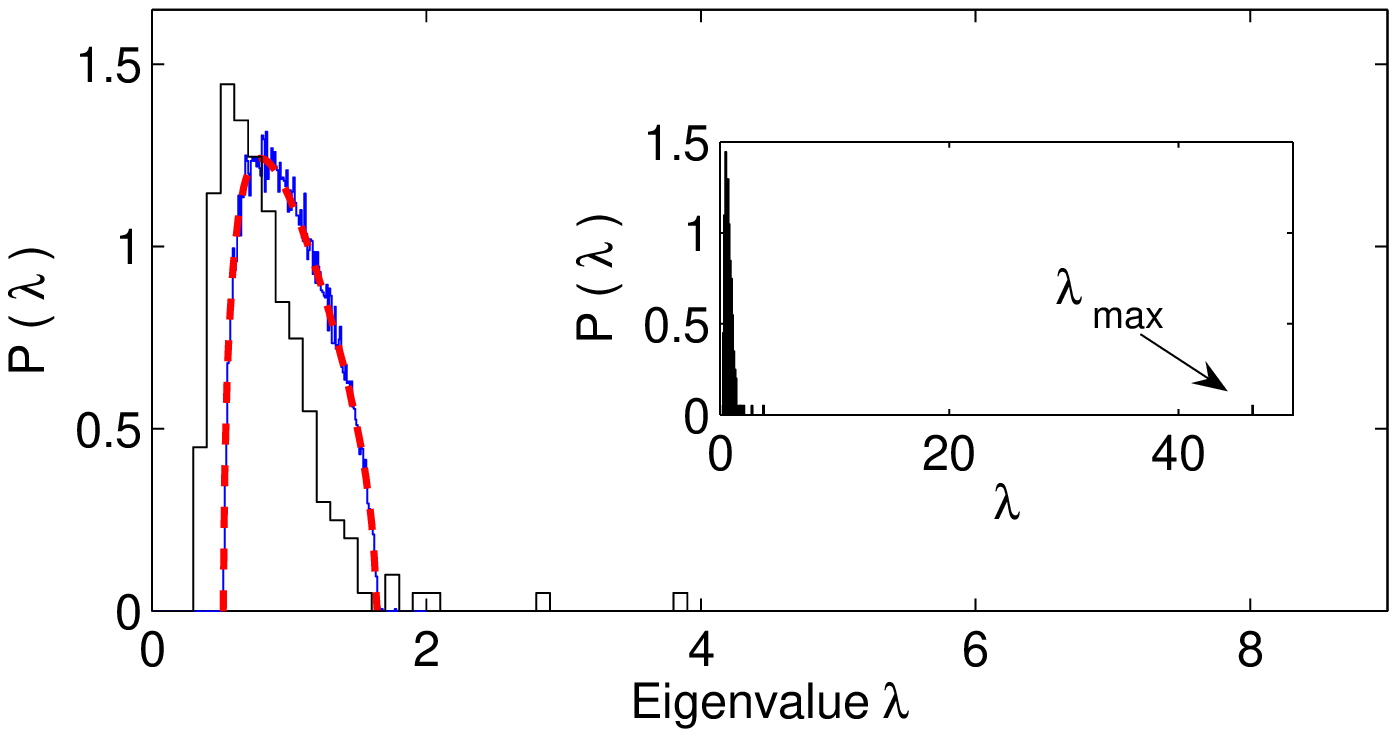}
\includegraphics[width=0.85\linewidth]{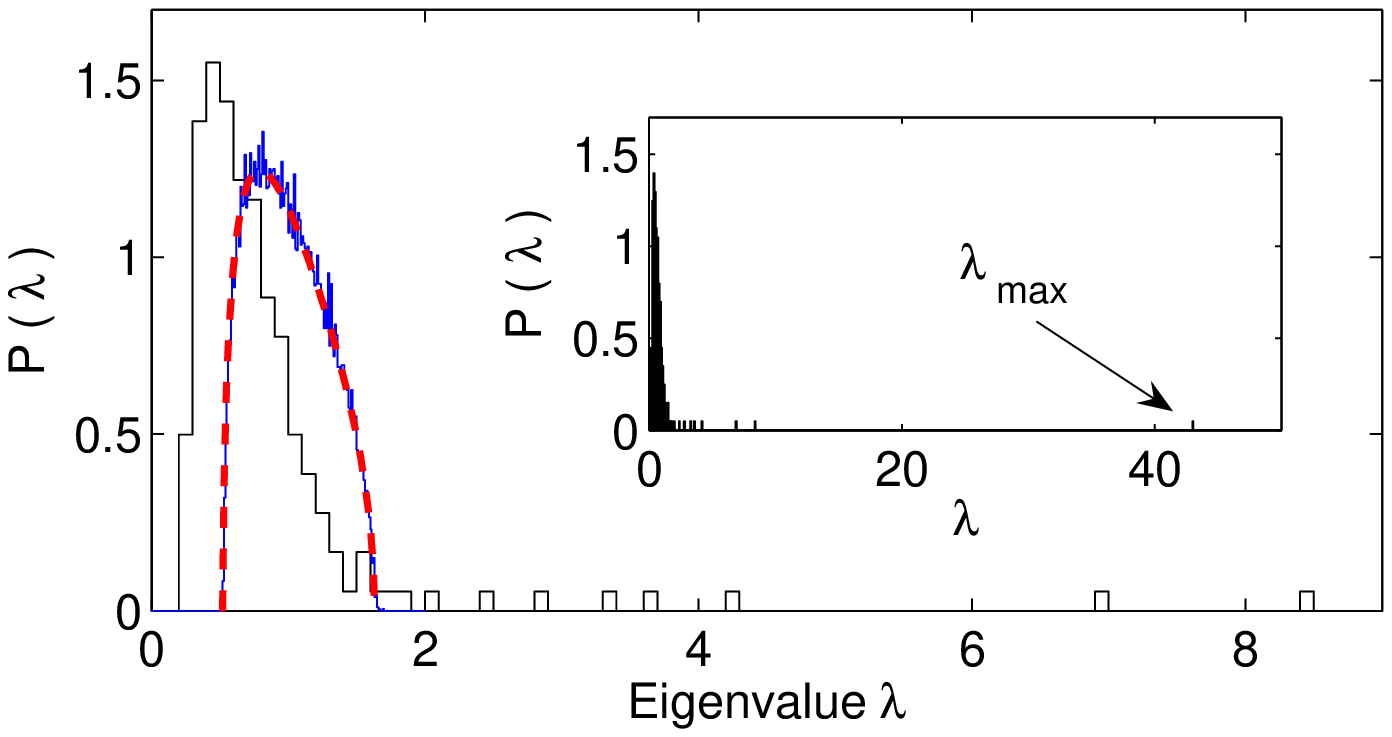}
\caption{The probability density function of the eigenvalues of the
correlation matrix ${\mathbf C}$ for NSE (top) and NYSE (bottom). For
comparison, the theoretical distribution predicted by
Eq.~(\ref{eq:sengupta}) is shown using broken curves, which overlaps with
the distribution obtained from the surrogate correlation matrix generated by
randomly shuffling each time series. In both figures, the inset shows the
largest eigenvalue.}
\label{fig:rmt}
\end{figure}

If the $N$ return time series of length $T$ are mutually uncorrelated, then
the resulting random correlation matrix is called a Wishart matrix,
whose statistical properties are well known~\cite{Sengupta99}. In the limit
$N\rightarrow\infty$, $T\rightarrow\infty$, such that $Q\equiv T/N\geq1$,
the eigenvalue distribution of this random correlation matrix is given by
\begin{equation} 
P_{\rm rm}( \lambda ) = \frac{Q}{2 \pi}
\frac{\sqrt{(\lambda_{max} - \lambda)(\lambda - \lambda_{min})}}{\lambda},
\label{eq:sengupta} 
\end{equation} 
for $\lambda_{min} \leq \lambda \leq \lambda_{max}$ and, $0$ otherwise. The
bounds of the distribution are given by $\lambda_{max,min} =
[1\pm(1/\sqrt{Q})]^2$. We now compare this with the statistical properties
of the empirical correlation matrix for the NSE. In the NSE data, there are
$N=201$ stocks each containing $T=2606$ returns; as a result $Q=12.97$.
Therefore, it follows that, in the absence of any correlation among the
stocks, the distribution should be bounded between $\lambda_{min}=0.52$ and
$\lambda_{max}=1.63$. As observed in developed
markets~\cite{Laloux99,Plerou99_prl,Plerou02,Utsugi04}, the bulk of the
eigenvalue spectrum $P(\lambda)$ for the empirical correlation matrix is in
agreement with the properties of a random correlation matrix spectrum
$P_{\rm rm}(\lambda)$, but a few of the largest eigenvalues deviate
significantly from the RMT bound (Fig.~\ref{fig:rmt}). However, the number
of these deviating eigenvalues are relatively few for NSE compared to NYSE.
To verify that these outliers are not an artifact of the finite length of
the observation period, we have randomly shuffled the return time series
for each stock, and then re-calculated the resulting correlation matrix.
The eigenvalue distribution for this surrogate matrix matches exactly with
the random matrix spectrum $P_{\rm rm}(\lambda)$, indicating that the
outliers are not due to ``measurement noise'' but are genuine indicators of
correlated movement among the stocks. Therefore, by analyzing the deviating
eigenvalues, we may be able to obtain an understanding of the structure of
interactions between the stocks in the market.

\subsection{Properties of the ``deviating'' eigenvalues}
\begin{figure*}
\centering
\includegraphics[width=0.85\linewidth]{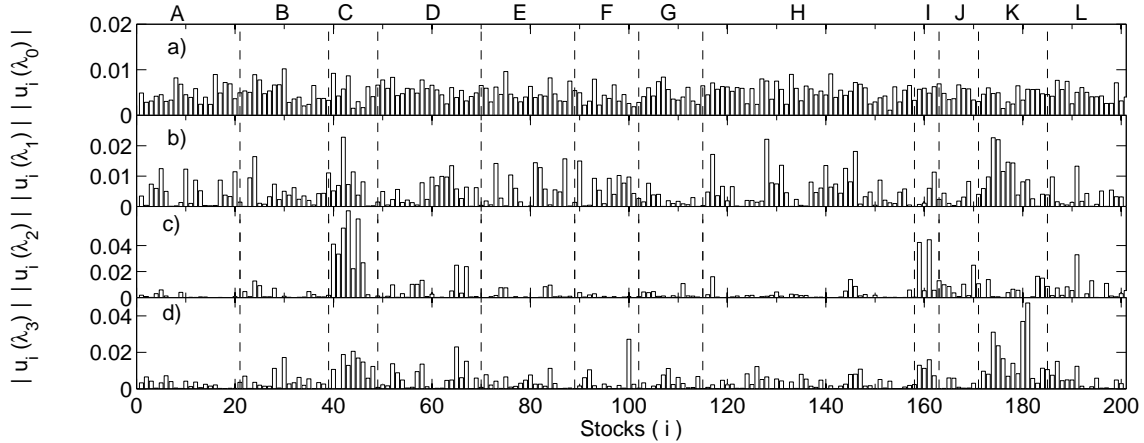}
\caption{The absolute values of the eigenvector components $u_i (\lambda)$
of stock $i$ corresponding to the first four largest eigenvalues of
${\mathbf C}$ for NSE. 
The stocks $i$ are arranged by business sectors separated by broken
lines. A: Automobile \& transport, B: Financial, C: Technology D: Energy,
E: Basic materials, F: Consumer goods, G: Consumer discretionary, H:
Industrial, I: IT \& Telecom, J: Services, K: Healthcare \& Pharmaceutical,
L: Miscellaneous.}
\label{fig:eigenvector} 
\end{figure*}

The largest eigenvalue $\lambda_0$ for the NSE cross-correlation matrix is
more than 28 times greater than the maximum predicted by RMT. This is
comparable to NYSE, where $\lambda_0$ is about 26 times 
greater than the
random matrix upper bound. Upon testing with synthetic US data containing
same number of missing data points as in the Indian market, we observed
that $\lambda_0$ remains almost unchanged compared to the value obtained
from the original US data.
%(mean and s.d. computed over 100 different realizations of 
%the synthetic data set), which is consistent with the
%result obtained from the original data.
The corresponding eigenvector shows a relatively
uniform composition, with all stocks contributing to it and all elements
having the same sign (Fig.~\ref{fig:eigenvector}, top). As this is
indicative of a common factor that affects all the stocks with the same
bias, the largest eigenvalue is associated with the {\em market mode},
i.e., the collective response of the entire market to external
information~\cite{Laloux99,Plerou02}. Of more interest for understanding
the market structure are the intermediate eigenvalues, i.e., those
occurring between the largest eigenvalue and the bulk of the distribution
predicted by RMT. For the NYSE, it was shown that corresponding
eigenvectors of these eigenvalues are localized, i.e., only a small number
of stocks, belonging to similar or related businesses, contribute
significantly to each of these modes~\cite{Gopikrishnan01,Plerou02}.
%From the synthetic US market data, we obtained the number of deviating
%eigenvalues to be $11.8 \pm 0.9$ (mean and s.d. computed over 100 different
%realizations of the synthetic data set), which is consistent with the
%observation of 11 deviating eigenvalues in the original data.
However, for NSE, although the Technology and the IT \& Telecom stocks are
dominant contributors to the eigenvector corresponding to the third
largest eigenvalue, a direct inspection of eigenvector composition does not
yield a straightforward interpretation in terms of a related group of
stocks corresponding to any particular eigenvalue
(Fig.~\ref{fig:eigenvector}).
%If we arrange stocks according to their corresponding business sectors, 
%we observe that the eigenvector component for the largest eigenvalue gets 
%equal contribution from all the sectors.

\begin{figure} 
\includegraphics[width=0.85\linewidth]{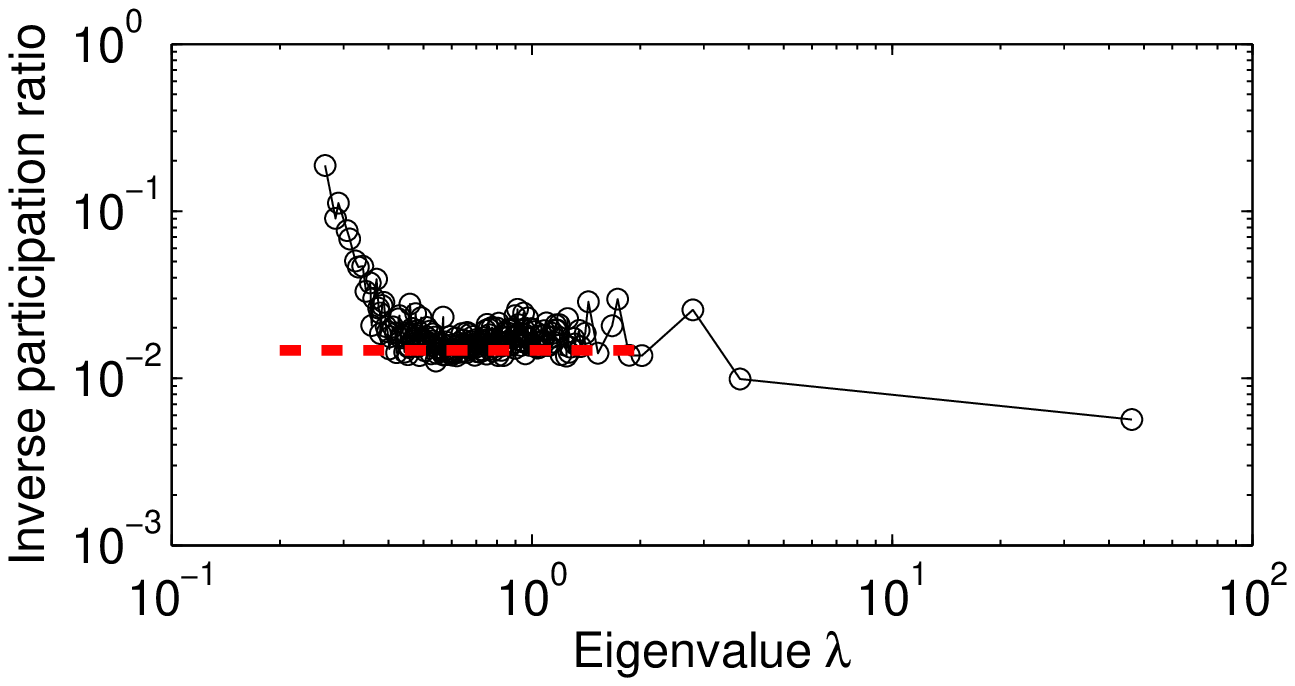}
\includegraphics[width=0.85\linewidth]{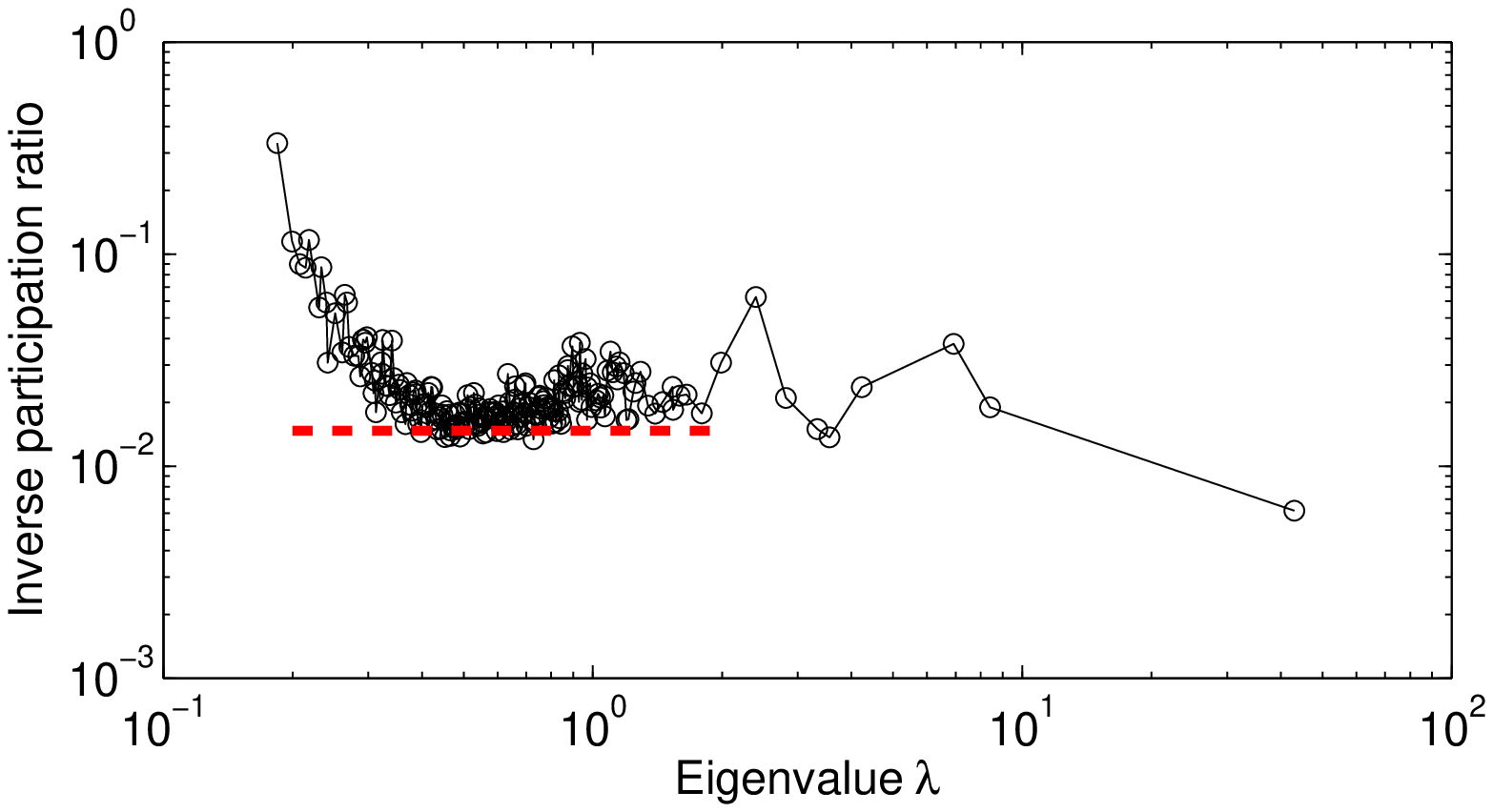}
\caption{Inverse participation ratio as a function of eigenvalue for the
correlation matrix ${\mathbf C}$ of NSE (top) and NYSE (bottom).  The
broken line indicates the average value of $\langle I \rangle = 1.49 \times
10^{-2}$ for the eigenvectors of a matrix constructed by randomly shuffling
each of the $N$ time series.}
\label{fig:ipr} 
\end{figure}

To obtain a quantitative measure of the number of stocks contributing to a
given eigenmode, we calculate the inverse participation ratio (IPR),
defined for the $k$th eigenvector as $I_k\equiv\sum_{i=1}^N [ u_{ki}]^4$,
where $u_{ki}$ are the components of eigenvector $k$. An eigenvector having
components with equal value, i.e., $u_{ki}=1/\sqrt{N}$ for all $i$, has 
$I_{k} = 1/N$. We
find this to be approximately true for the eigenvector corresponding to the
largest eigenvalue, which represents the market mode. To see how different
stocks contribute to the remaining eigenvectors, we note that if a single
stock had a dominant contribution in any eigenvector, e.g., $u_{k1}=1$ and
$u_{ki}=0$ for $i \neq 1$, then $I_{k}=1$ for that eigenvector. Thus, IPR
gives the reciprocal of the number of eigenvector components (and
therefore, stocks) with significant contribution. On the other hand, the
average value of $I_{k}$, for eigenvectors of a random correlation matrix
obtained by randomly shuffling the time series of each stock, is $\langle I
\rangle = 3/N \approx 1.49 \times 10^{-2}$. Fig.~\ref{fig:ipr} shows that the
eigenvalues belonging to the bulk of the spectrum indeed have 
this value of IPR. But at the lower and higher end of eigenvalues, both the
US and Indian markets show deviations, suggesting the existence of
localized modes. However, these deviations are much less significant and
fewer in number in the latter compared to the former. This implies that
distinct groups, whose members are mutually correlated in their price
movement, do exist in NYSE, while their existence is far less clear in NSE.

\subsection{Filtering the correlation matrix}
\begin{figure}
\includegraphics[width=0.85\linewidth]{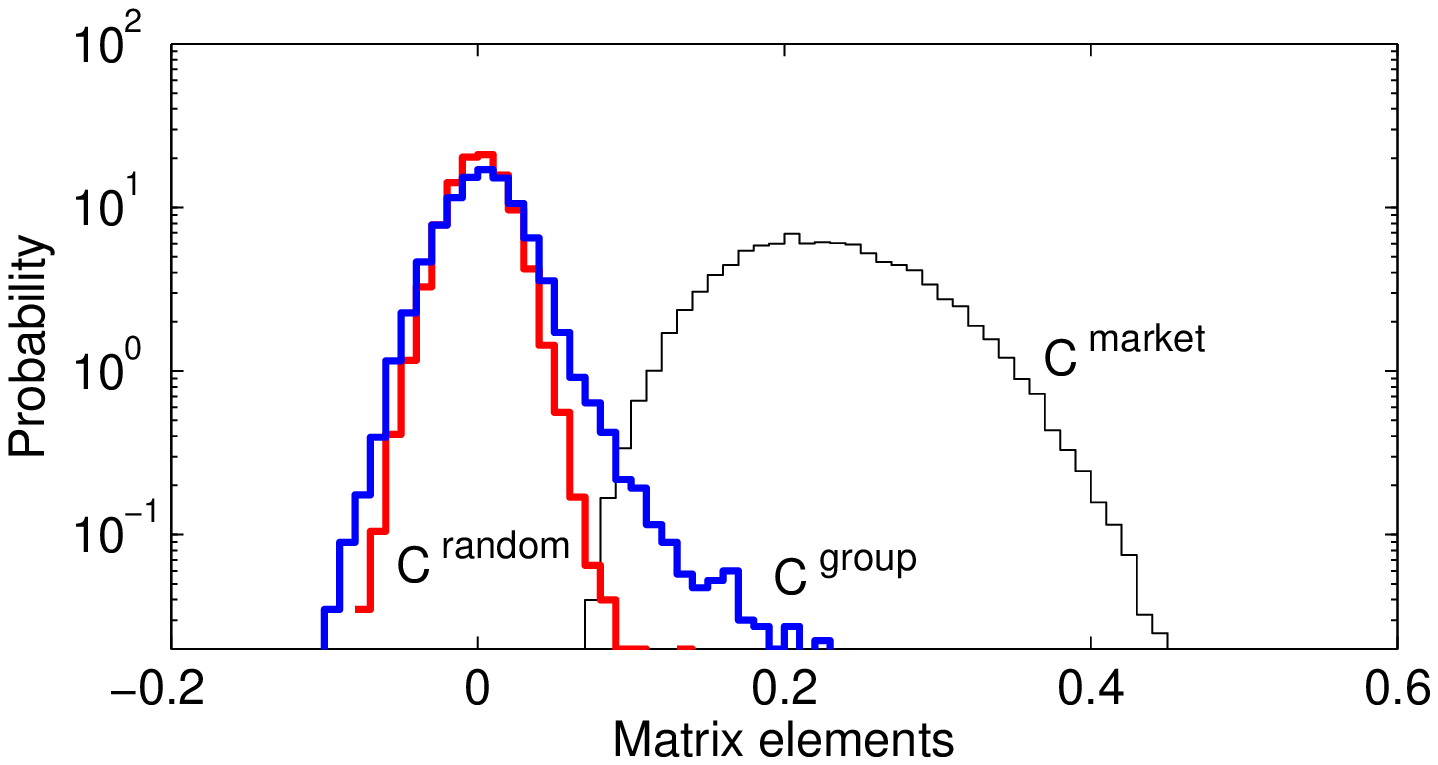}
\includegraphics[width=0.85\linewidth]{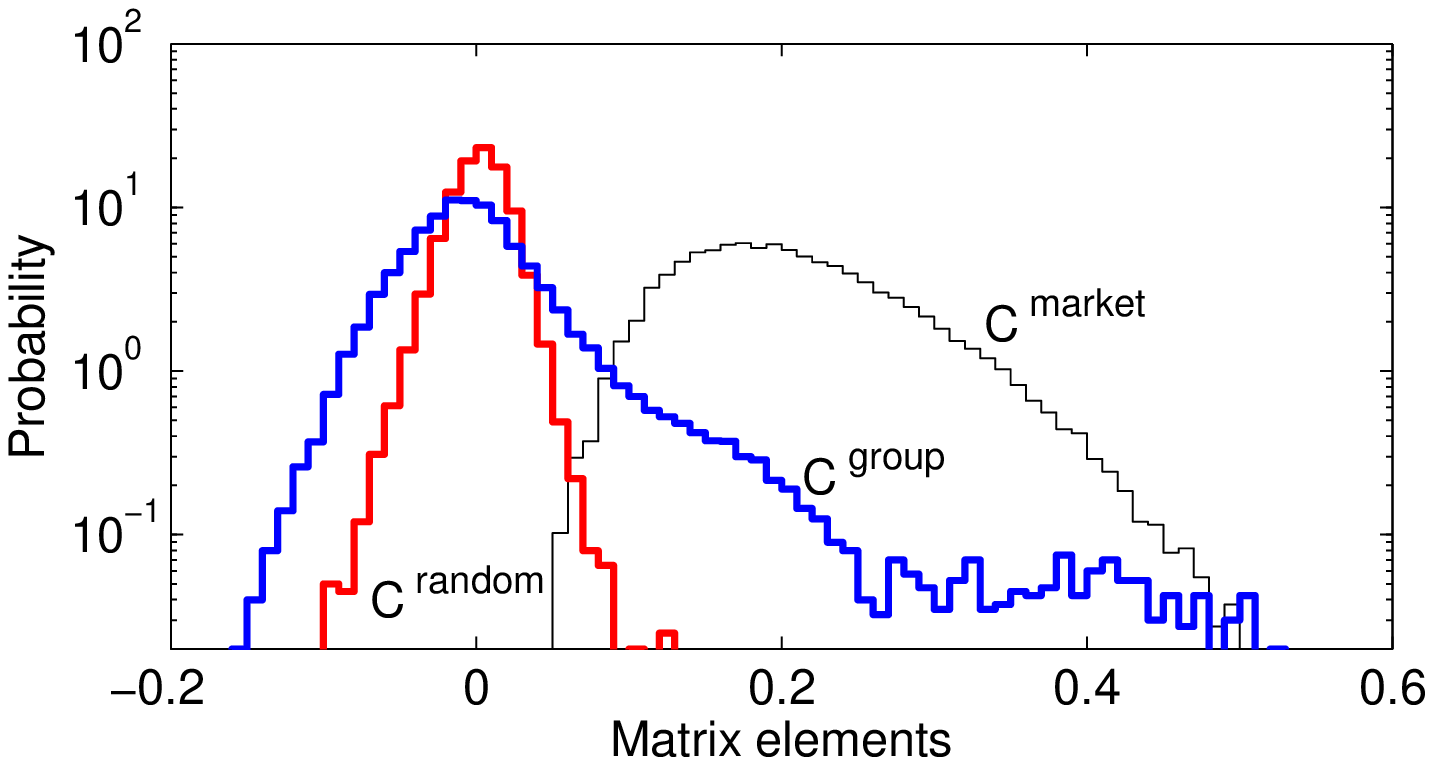}
\caption{The distribution of elements of correlation matrix corresponding
to the market, ${\mathbf C}^{market}$, the group, ${\mathbf C}^{group}$, and the
random interaction, ${\mathbf C}^{random}$. For NSE (top) $N_{g}=5$ whereas for
NYSE (bottom) $N_{g}=10$. The short tail for the distribution of the
${\mathbf C}^{group}$ elements in NSE indicates that the correlation generated by mutual
interaction among stocks is relatively weak.}
\label{fig:compare}
\end{figure}

The above analysis suggests the existence of a market-induced
correlation across all stocks, which makes it difficult to observe the
correlations that might be due to interactions between stocks belonging to
the same sector. Therefore, we now use a filtering method to remove 
%for identifying the different correlated stock groups by removing the 
market mode, as well as the random noise~\cite{Kim05}. The correlation
matrix is first decomposed as 
\begin{equation}
{\mathbf C}= \sum_{i=0}^{N-1}\lambda_i{\mathbf u}_i{\mathbf u}_i^T,
\end{equation}
where $\lambda_{i}$ are the eigenvalues of ${\mathbf C}$ sorted in
descending order and ${\mathbf u}_i$ are corresponding eigenvectors. As
only the eigenvectors corresponding to the few largest eigenvalues are
believed to contain information on significantly correlated stock groups,
the contribution of the intra-group correlations to the ${\mathbf C}$ matrix
can be written as a partial sum of $\lambda_{\alpha} {\mathbf
u}_{\alpha}{\mathbf u}_{\alpha}^T$, where $\alpha$ is the index of the
corresponding eigenvalue. Thus, the correlation matrix can be decomposed
into three parts, corresponding to the {\em market}, {\em group} and {\em
random} components:
\begin{eqnarray}
{\mathbf C} &=& {\mathbf C}^{market} + {\mathbf C}^{group} + {\mathbf
C}^{random} \nonumber \\ 
 &=& \lambda_0 {\mathbf u}_0 {\mathbf u}_0^T + 
 \sum_{i =1}^{N_{g}}\lambda_{i} {\mathbf u}_i {\mathbf u}_i^T +
 \sum_{i = N_{g}+1}^{N-1}\lambda_{i} {\mathbf u}_i {\mathbf u}_i^T,
\end{eqnarray}
where, $N_{g}$ is the number of eigenvalues (other than the largest one)
which deviates from the bulk of the eigenvalue spectrum. For NSE we have
chosen $N_{g}=5$. However, the exact value of this choice is not crucial as
small changes in $N_{g}$ do not alter the results, the error involved being
limited to the eigenvalues closest to the bulk that have the smallest
contribution to ${\mathbf C}^{group}$. Fig.~\ref{fig:compare} shows the
result of decomposing the correlation matrix into the three components, for
both the Indian and US markets. Compared to the latter, the distribution of
matrix elements of ${\mathbf C}^{group}$ in the former shows a
significantly truncated tail. This indicates that intra-group correlations
are not prominent in NSE, whereas they are comparable with the overall
market correlations in NYSE. It follows that the collective behavior in the
Indian market is dominated by external information that affects all stocks.
Correspondingly, correlations generated by interactions between stocks, as
would be the case for stocks in a given business sector, are much weaker,
and hence, such correlated sectors would be difficult to observe.

\begin{figure}
\includegraphics[width=0.85\linewidth]{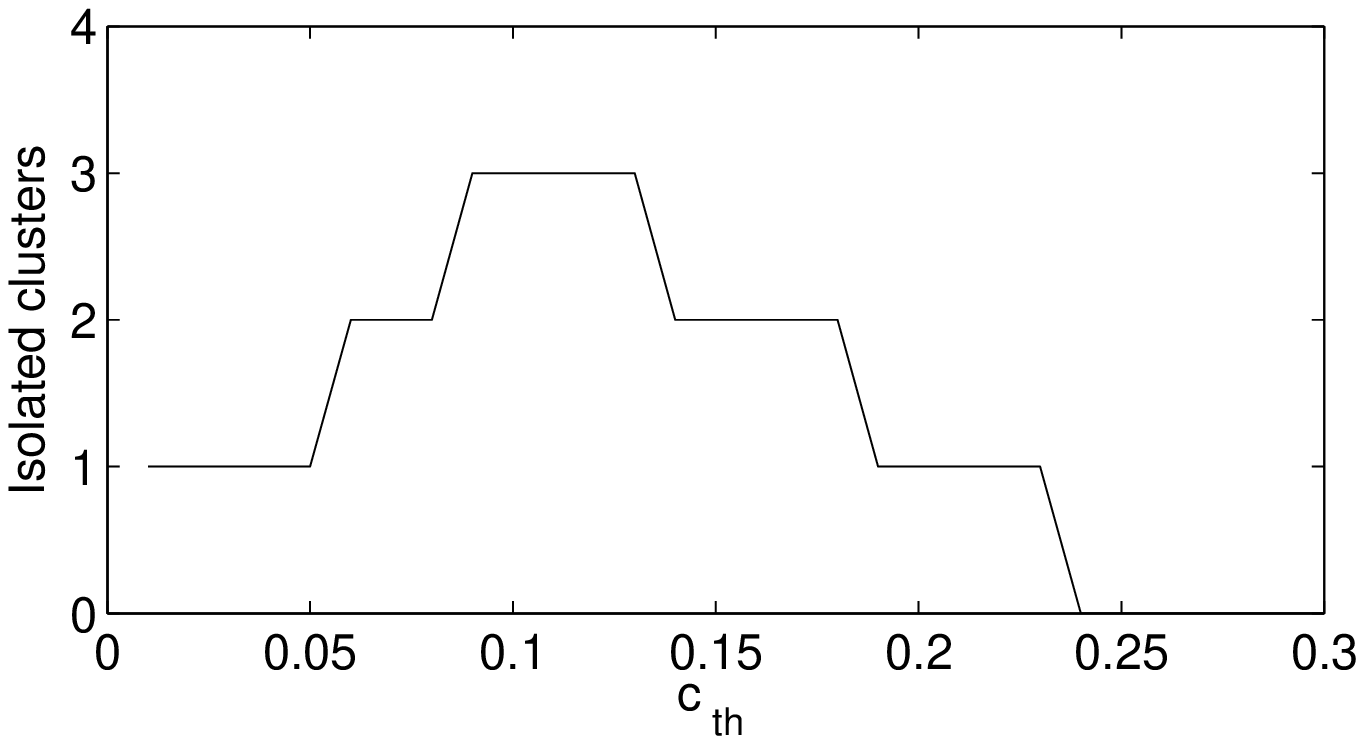}
\caption{The number of isolated clusters in the interaction network for 
NSE stocks as a function of the threshold value $c_{th}$. For low $c_{th}$ the
network consist of a single cluster which contains all the nodes, whereas
for high $c_{th}$ the network consists only of isolated nodes.}
\label{fig:cstar}
\end{figure}
\begin{figure*} 
\includegraphics[width=0.8\linewidth]{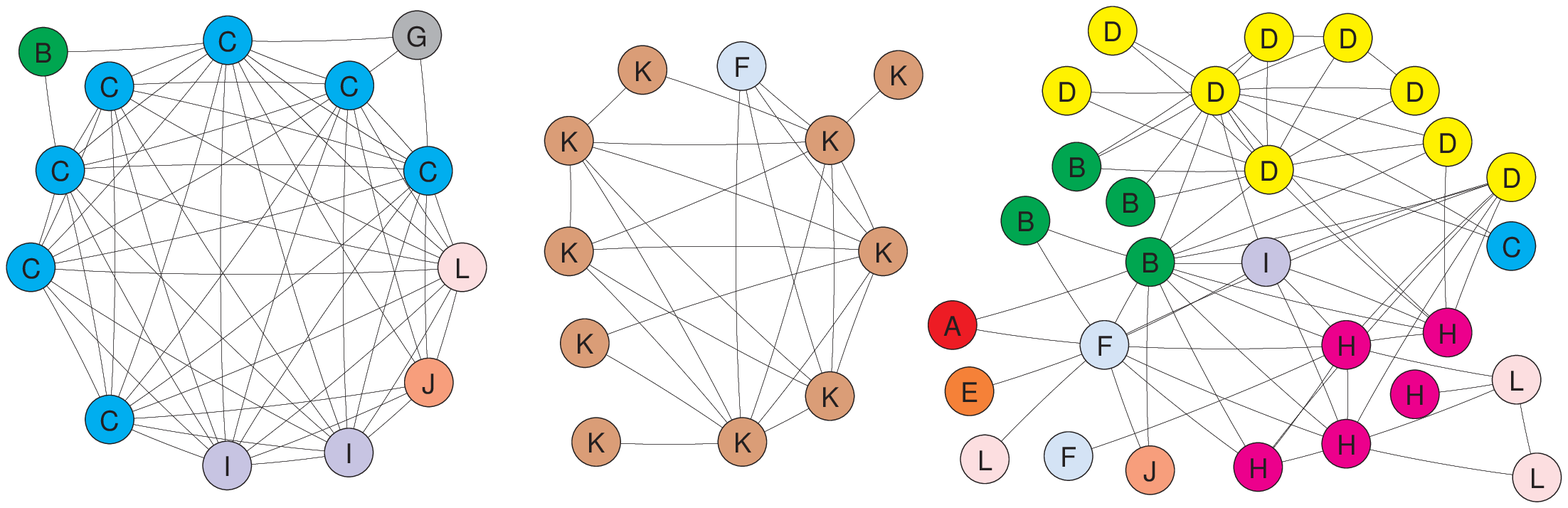}
\caption{The structure of interaction network in the Indian financial
market at threshold $c^* = 0.09$. The left cluster
comprises of mostly Technology stocks, while the middle cluster is
composed almost entirely of Healthcare \& Pharmaceutical stocks. By
contrast, the cluster on the right is not dominated by any particular
sector. The node labels indicate the business sector to which a stock
belongs and are as specified in the caption to Fig~\ref{fig:eigenvector}.}
%A: Automobile \& transport (red), B: Financial (green), C:Technology
%(cyan) D: Energy (yellow), E: Basic materials (orange), F: Consumer goods
%(lskyblue), G: Consumer discretionary (gray), H: Industrial (magenta), I:
%IT \& Telecom (lightpurple), J: Services (melon), K: Healthcare \&
%Pharmaceutical (tan), L: Miscellaneous (pink).
%The figure has been drawn using the Pajek software.
\label{fig:network} 
\end{figure*}

We indeed find this to be true when we use the information in the group
correlation matrix to construct the network of interacting
stocks~\cite{Kim05}. The adjacency matrix ${\mathbf A}$ of this network is
generated from the group correlation matrix ${\mathbf C}^{group}$ by using
a threshold $c_{th}$ such that $A_{ij} = 1$ if $C_{ij}^{group} > c_{th}$,
and $A_{ij} = 0$ otherwise. Thus, a pair of stocks are connected if the
group correlation coefficient $C_{ij}^{group}$ is larger than a preassigned
threshold value, $c_{th}$. To determine an appropriate choice of
$c_{th}=c^{*}$ we observe the number of isolated clusters (a cluster
being defined as a group of connected nodes) in the network
for a given $c_{th}$ (Fig.~\ref{fig:cstar}). We found this number to be
much less in NSE compared to that observed in NYSE for any value of
$c_{th}$~\cite{Kim05}. Fig.~\ref{fig:network} shows the resultant network
for $c^* = 0.09$, for which the largest number of isolated clusters of
stocks are obtained. The network has 52 nodes and 298 links partitioned
into 3 isolated clusters. From these clusters, only two business sectors
can be properly identified, namely the Technology and the Pharmaceutical
sectors. The fact that the majority of the NSE stocks cannot be arranged
into well-segregated groups reflecting business sectors illustrates our 
conclusion that intra-group interaction is much weaker than the
market-wide correlation in the Indian market.

\subsection{Relating correlation with market evolution}
\begin{figure}
\centering
\includegraphics[width=0.85\linewidth]{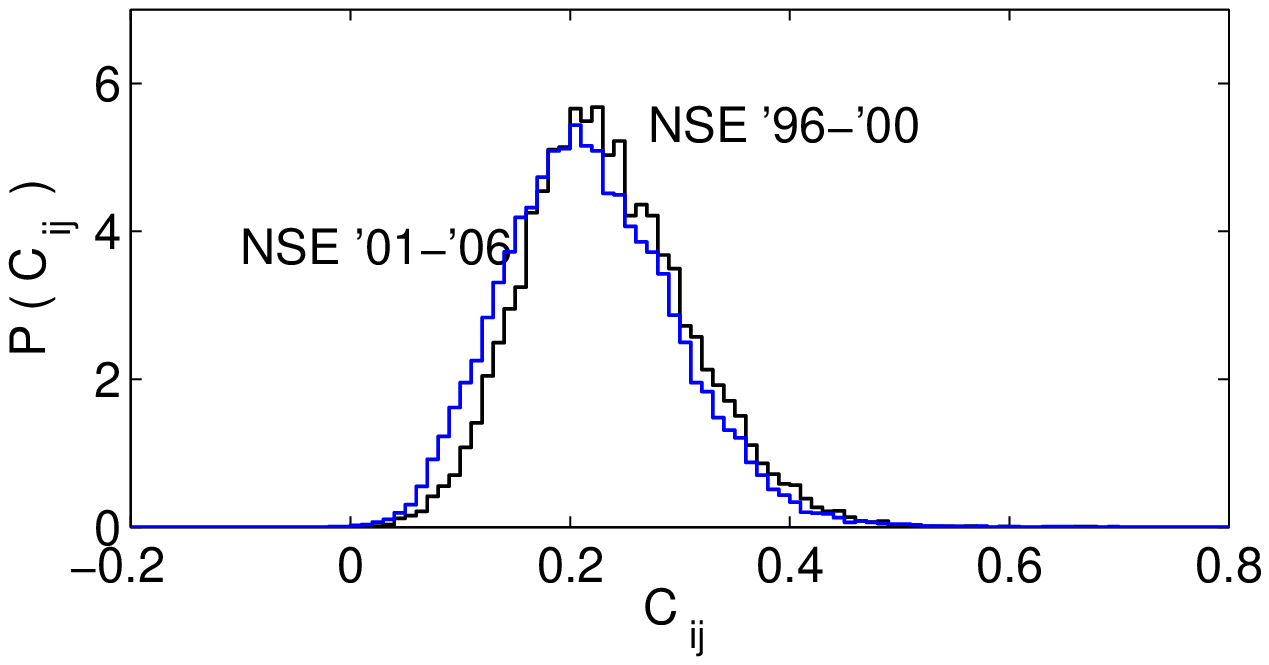}
\caption{The probability density functions of the elements in the
correlation matrix ${\mathbf C}$ for NSE
during (a) the period Jan 1996-Dec 2000 and (b) Jan 2001-May 2006. 
The mean value of the elements of
$\mathbf C$ for the two periods are 0.23 and 0.21,
respectively.}
\label{fig:comparison1}
\end{figure}

We now compare between two different time intervals in the NSE data.
For convenience we divide the data set into two non-overlapping parts
corresponding to the periods between Jan 1996-Dec 2000 (Period I) and between
Jan 2001-May 2006 (Period II). The corresponding correlation 
matrices ${\mathbf C}$ are
generated following the same set of steps as for the entire data set.
The average value for the elements of the correlation matrix is slightly
lower for the later period, suggesting a greater homogeneity
between the stocks at the earlier period (Fig.~\ref{fig:comparison1}).

\begin{figure}
\centering
\includegraphics[width=0.85\linewidth]{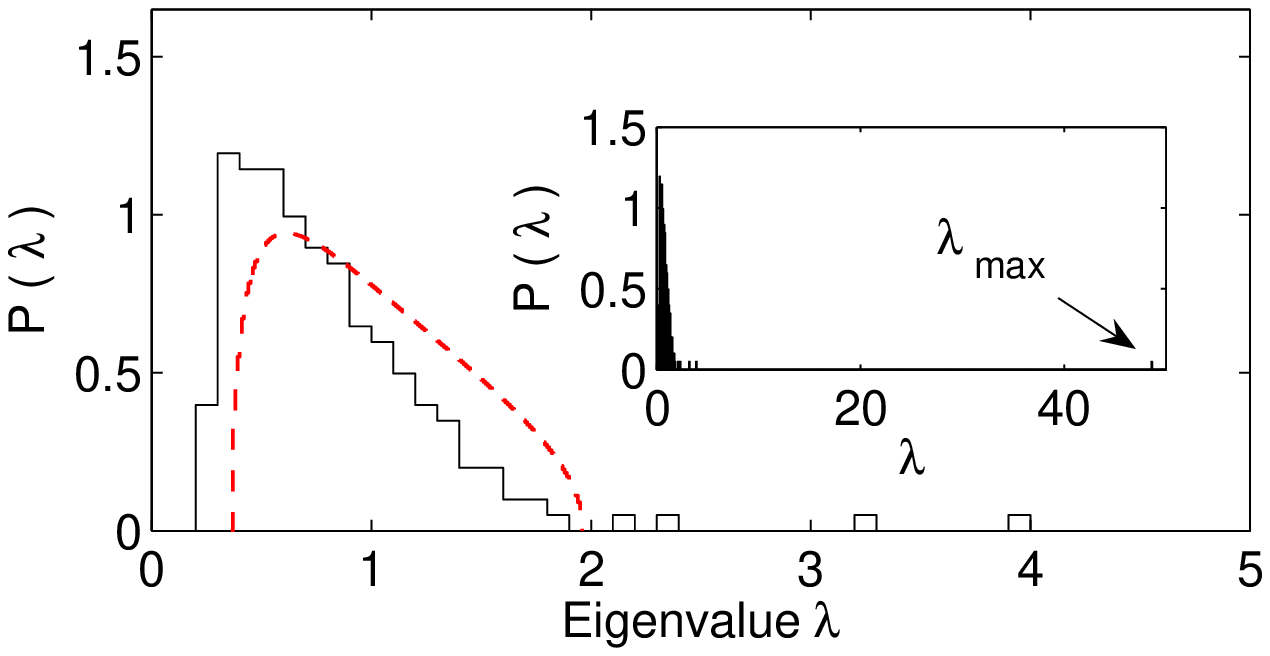}
\includegraphics[width=0.85\linewidth]{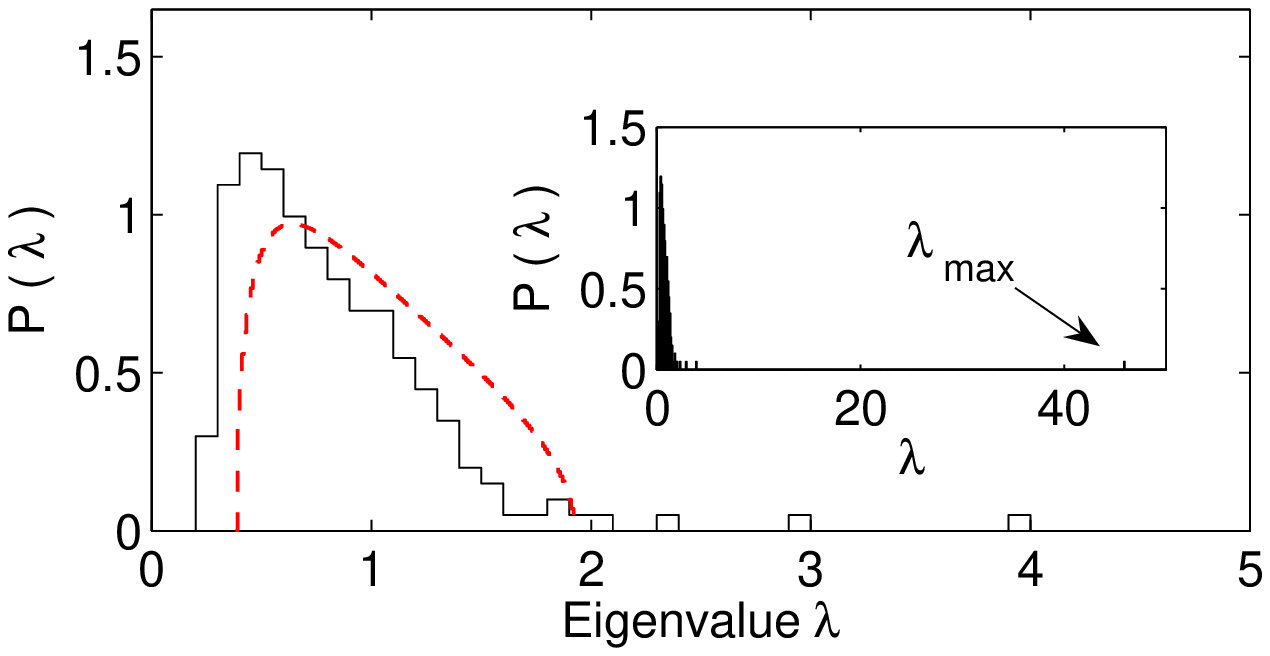}
\caption{The probability density function of the eigenvalues of the
NSE correlation matrix ${\mathbf C}$ for the periods (top) Jan 1996-Dec 2000
and (bottom) Jan 2001-May 2006. For
comparison, the theoretical distribution predicted by
Eq.~(\ref{eq:sengupta}) is shown using broken curves.
In both figures, the inset shows the largest eigenvalue.}
\label{fig:comparison2}
\end{figure}

Next, we look at the eigenvalue distribution of ${\mathbf C}$ for the
two periods (Fig.~\ref{fig:comparison2}. The Q value for Period I is 6.21, while for
Period II it is 6.77. Thus the bounds for the
random distribution is almost same in the two cases. In contrast, the
largest deviating eigenvalues, $\lambda_0$, are
different: 48.56 for Period I and 45.88 for Period II.
This implies the relative dominance of the market mode in the earlier 
period, again suggesting that with time the market has become less
homogeneous.
The number of deviating eigenvalues remain the same for the two periods.

When the interaction networks between stocks are generated for the two periods,
they show less distinction into clearly defined sectors than was obtained with 
the data for the entire period. 
This is possibly because the shorter data sets 
create larger fluctuations in the correlation values, thereby making
it difficult to segregate the existing market sectors. However, we do observe
that, using the same threshold value for generating networks in the two 
periods yield, for the later period, isolated clusters that are distinguishable into distinct sub-clusters connected to each other via a few links only, 
whereas in the earlier period the clusters are much more homogeneous. 
This implies that as the Indian market is evolving, the interactions between
stocks are tending to get arranged into clearly identifiable groups.
We propose that such structural re-arrangement in the interactions
is a hallmark of emerging markets as they evolve into developed ones.
\section{Model of Market Dynamics}
\label{sec:4}
To understand the relation between the interaction structure among stocks
and the eigenvalues of the correlation matrix, we perform a multivariate
time series analysis using a simple two-factor model of market dynamics. We
assume that the normalized return at time $t$ of the $i$th stock from the
$k$th business sector can be decomposed into (i) a market factor $r_m(t)$,
that contains information or signal common to all stocks, (ii) a sector
factor $r_g^{k}(t)$, representing effects exclusive to stocks in the $k$th
sector, and (iii) an idiosyncratic term, $\eta_{i}(t)$, which corresponds
to random variations unique for that stock. Thus, 
\begin{equation}
r_{i}^{k}(t) = \beta_{i} r_m(t)+\gamma_{i}^{k} r_g^{k}(t)+\sigma_{i} \eta_{i}(t),
\label{modeleqn1}
\end{equation}
where $\beta_i$, $\gamma_i^k$ and $\sigma_i$ represent relative strengths
of the three terms mentioned above, respectively. For simplicity, these
strengths are assumed to be time independent.  We choose
$r_m(t)$, $r_g^{k}(t)$ and $\eta_{i}(t)$ from a zero mean and unit variance
Gaussian distribution. We further assume that the normalized returns
$r_{i}$, also follow Gaussian distribution with zero mean and unit
variance. Although the empirically observed return distributions have power
law tails, as these distributions are not Levy stable, they will converge
to Gaussian if the returns are calculated over sufficiently long intervals.
The assumption of unit variance for the returns ensures that the relative 
strengths
of the three terms will follow the relation:
\begin{equation}
{\beta_{i}}^2 + ({\gamma_{i}^k})^2 + {\sigma_{i}}^2 = 1. 
\label{eq:modeleqn2}
\end{equation}
As a result, for each stock we can assign $\sigma_i$ and $\gamma_i$ 
independently, and obtain $\beta_i$ from Eq.~(\ref{eq:modeleqn2}). 
We choose $\sigma_i$ and $\gamma_i$ from a uniform distribution having
width $\delta$ and centered about the mean values $\sigma$ and $\gamma$,
respectively.

\begin{figure}
\includegraphics[width=0.9\linewidth]{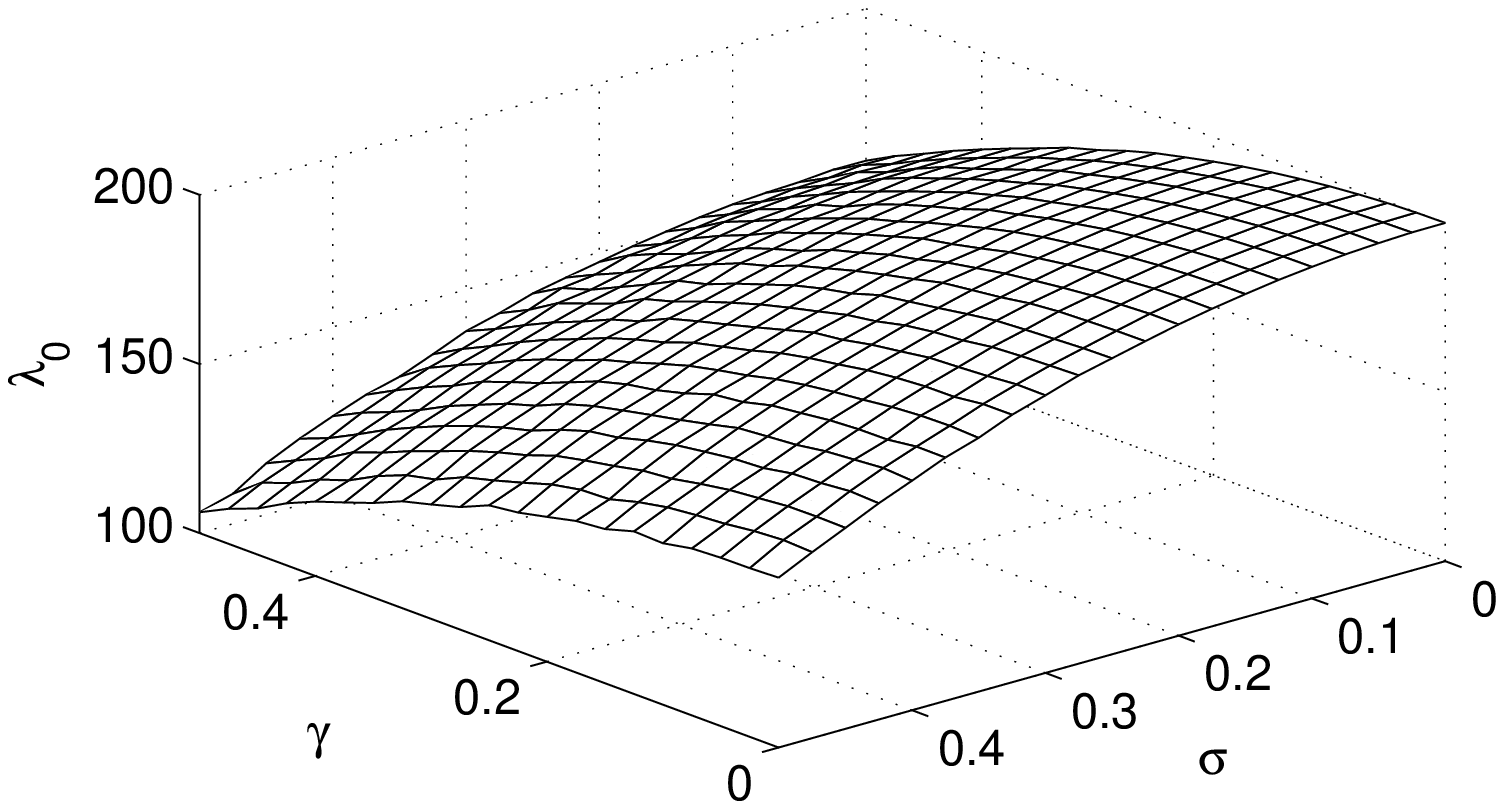}
\includegraphics[width=0.9\linewidth]{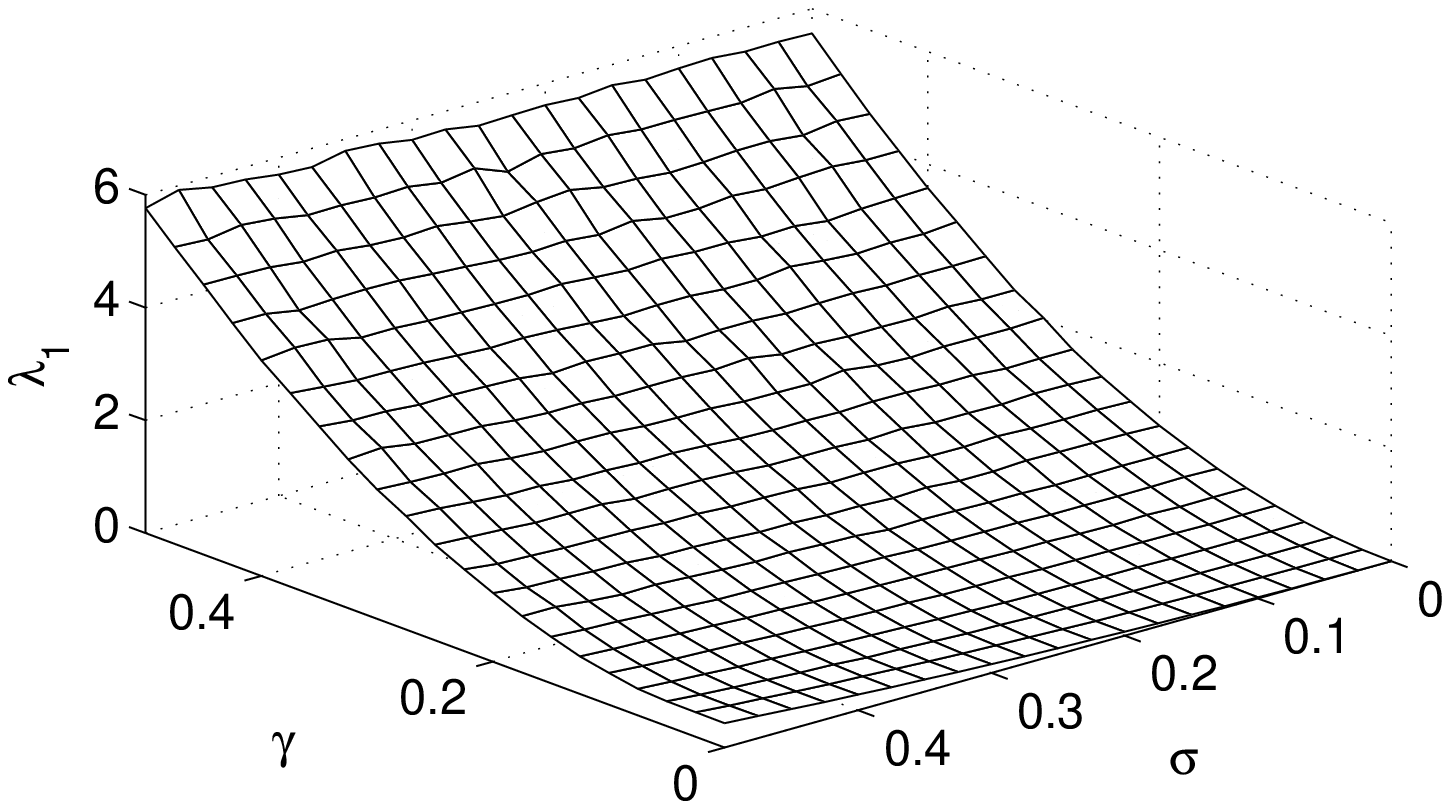}
\caption{The variation of the largest (top) and second largest (bottom)
eigenvalues of the correlation matrix of simulated return in the
two-factor model (Eq.~\ref{modeleqn1}) with the model parameters $\gamma$
and $\sigma$ (corresponding to strength of the sector and idiosyncratic
effects, respectively). The matrix is constructed for $N = 200$ stocks
each with return time series of length $T = 2000$ days. We assume there to
be 10 sectors, each having 20 stocks.}
\label{fig:factormodel}
\end{figure}
We now simulate an artificial market with $N$ stocks belonging to $K$
sectors by generating time series of length $T$ for returns $r_i^k$ from
the above model. These $K$ sectors are composed of
$n_{1},n_{2},\dots,n_{K}$ stocks such that $n_{1}+n_{2}+\dots+n_{K}=N$.
The collective behavior is then analysed by constructing the resultant
correlation matrix ${\mathbf C}$ and obtaining its eigenvalues. Our aim is to
relate the spectral properties of ${\mathbf C}$ with the underlying structure of
the market given by the relative strength of the factors. We first
consider the simple case, where the contribution due to market factor is
neglected, i.e., $\beta_{i}=0$ for all $i$, and the strength of sector factor
is equal for all stocks within a sector, i.e., $\gamma_i^k = \gamma^k$, is
independent of $i$. In this case, the spectrum of the correlation matrix
is composed of $K$ large eigenvalues, $1+(n_{j}-1)(\gamma^j)^2$, where $j=1\dots
K$, and $N-K$ small eigenvalues, $1-(\gamma^j)^2$, each with degeneracy
$n_{j}-1$, where $j=1\dots K$~\cite{Lillo05}. Now, we consider nonzero
market factor which is equal for all stocks i.e., 
$\beta_{i}=\beta$ for all $i$, and the 
strength of sector factor is also same for all stocks, i.e.,
$\gamma_i^k = \gamma$ (independent of $i$ and $k$). In this case too, there are
$K$ large eigenvalues and $N-K$ small eigenvalues. Our numerical simulations
suggest that the largest and the second largest eigenvalues are
\begin{eqnarray}
\lambda_0 &\sim& N \beta^2, \nonumber \\ 
\lambda_1 &\sim& n_{l} ( 1 - \beta^2),
\end{eqnarray}
respectively, where $n_{l}$ is the size of the largest sector, while 
the $N-K$ small
degenerate eigenvalues are $1-\beta^2-\gamma^2$. We now choose the
strength $\gamma_{i}^k$ and $\sigma_{i}$ from a uniform
distribution with mean $\gamma$ and $\sigma$ respectively and with width
$\delta = 0.05$. Fig.~\ref{fig:factormodel} shows the variation of the
largest and second largest eigenvalues with $\sigma$ and $\gamma$. The
strength of the market factor is determined from Eq.\ref{eq:modeleqn2}.

Note that, decreasing the strength of the sector factor relative to the
market factor results in decreasing the second largest eigenvalue
$\lambda_1$. As $Q = T/N$ is fixed, the RMT bounds for the bulk of the
eigenvalue distribution, $[\lambda_{min},\lambda_{max}]$, remain unchanged.
Therefore, a decrease in $\lambda_1$ implies that the large intermediate
eigenvalues occur closer to the bulk of the spectrum predicted by RMT, as
is seen in the case of NSE. The analysis of the model supports our
hypothesis that the spectral properties of the correlation matrix for the
NSE are consistent with a market in which the effect of information common
for all stocks (i.e., the market mode) is dominant, resulting in all stocks
exhibiting a significant degree of correlation. 

\section{Conclusions}
\label{sec:5}
In conclusion, we demonstrate that the stocks in emerging market are much
more correlated than in developed markets. Although, the bulk of the
eigenvalue spectrum of the correlation matrix of stocks $\mathbf C$ in
emerging market is similar to that observed for developed markets, the
number of eigenvalues deviating from the RMT upper bound are smaller in number.
Further, most of the observed correlation among stocks is found to be due
to effects common to the entire market, whereas correlation due to
interaction between stocks belonging to the same business sector are weak.
This dominance of the market mode relative to modes arising through
interactions between stocks makes an emerging market appear more
correlated than developed markets. Using a simple two-factor model we show
that a dominant market factor, relative to the sector factor, results in
spectral properties similar to that observed empirically for the Indian
market. Our study helps in understanding the evolution of markets as
complex systems, suggesting that strong interactions may emerge within groups
of stocks as a market evolves over time. How such self-organization occurs
and its relation to other changes that a market undergoes during its
development, e.g., large increase in transaction volume, is a question
worth pursuing in the future with the tools available to econophysicists.

Our paper also makes a significant point regarding the physical
understanding of markets as complex dynamical systems. In recent times, the
role of the interaction structure within a market in governing its overall
dynamical properties has come under increasing scrutiny. However, such
intra-market interactions affect only very weakly certain market
properties, which is underlined by the observation of identical fluctuation
behaviour in markets having very different interaction structures, viz.,
NYSE and NSE~\cite{Pan07,Pan06}. The system can be considered as a single
homogeneous entity responding only to external signals in explaining these
statistical features, e.g., the price fluctuation distribution. This
suggests that the basic assumption behind the earlier approach of studying
financial markets as essentially executing random walks in response to
independent external shocks~\cite{Bachelier00}, which ignored the internal
structure, may still be considered to be accurate for explaining market
fluctuation phenomena. In other words, complex interacting systems like
financial markets can have simple mean field-like description for some of
their properties.

\acknowledgments{
We thank N.~Vishwanathan for assistance in preparing the data for analysis.
We also thank M.~Marsili and M.S.~Santhanam for helpful discussions.}

% BibTeX users please use
% \bibliographystyle{plain}
% \bibliographystyle{unsrt}

\bibliography{ecno_reference}

\begingroup
\squeezetable
\begin{table*}
	\caption{The list of 201 stocks of NSE analyzed in this paper.}
	\begin{ruledtabular}
		\begin{tabular}{cll |@{\mbox{ }} cll|@{\mbox{ }} cll}
			$i$ & Company & Sector & $i$ & Company & Sector & $i$ & Company & Sector\\ \hline
			1 & UCALFUEL   & Automobiles Transport &  68 & IBP	       & Energy                &  135 & HIMATSEIDE & Industrial     \\ 
			2 & MICO	      & Automobiles Transport &  69 & ESSAROIL   & Energy                &  136 & BOMDYEING  & Industrial     \\
			3 & SHANTIGEAR & Automobiles Transport &  70 & VESUVIUS   & Energy                &  137 & NAHAREXP   & Industrial     \\
			4 & LUMAXIND   & Automobiles Transport &  71 & NOCIL	     & Basic Materials       &  138 & MAHAVIRSPG & Industrial     \\
			5 & BAJAJAUTO  & Automobiles Transport &  72 & GOODLASNER & Basic Materials       &  139 & MARALOVER  & Industrial     \\
			6 & HEROHONDA  & Automobiles Transport &  73 & SPIC	     & Basic Materials       &  140 & GARDENSILK & Industrial     \\
			7 & MAHSCOOTER & Automobiles Transport &  74 & TIRUMALCHM & Basic Materials       &  141 & NAHARSPG   & Industrial     \\
			8 & ESCORTS    & Automobiles Transport &  75 & TATACHEM   & Basic Materials       &  142 & SRF	      & Industrial     \\
			9 & ASHOKLEY   & Automobiles Transport &  76 & GHCL	     & Basic Materials       &  143 & CENTENKA   & Industrial     \\
			10 & M\&M	      & Automobiles Transport &  77 & GUJALKALI  & Basic Materials       &  144 & GUJAMBCEM  & Industrial     \\
			11 & EICHERMOT  & Automobiles Transport &  78 & PIDILITIND & Basic Materials       &  145 & GRASIM     & Industrial     \\
			12 & HINDMOTOR  & Automobiles Transport &  79 & FOSECOIND  & Basic Materials       &  146 & ACC	      & Industrial     \\
			13 & PUNJABTRAC & Automobiles Transport &  80 & BASF	     & Basic Materials       &  147 & INDIACEM   & Industrial     \\
			14 & SWARAJMAZD & Automobiles Transport &  81 & NIPPONDENR & Basic Materials       &  148 & MADRASCEM  & Industrial     \\
			15 & SWARAJENG  & Automobiles Transport &  82 & LLOYDSTEEL & Basic Materials       &  149 & UNITECH    & Industrial     \\
			16 & LML        & Automobiles Transport &  83 & HINDALC0   & Basic Materials       &  150 & HINDSANIT  & Industrial     \\
			17 & VARUNSHIP  & Automobiles Transport &  84 & SAIL	     & Basic Materials       &  151 & MYSORECEM  & Industrial     \\
			18 & APOLLOTYRE & Automobiles Transport &  85 & TATAMETALI & Basic Materials       &  152 & HINDCONS   & Industrial     \\
			19 & CEAT       & Automobiles Transport &  86 & MAHSEAMLES & Basic Materials       &  153 & CARBORUNIV & Industrial     \\
			20 & GOETZEIND  & Automobiles Transport &  87 & SURYAROSNI & Basic Materials       &  154 & SUPREMEIND & Industrial     \\
			21 & MRF	      & Automobiles Transport &  88 & BILT	     & Basic Materials       &  155 & RUCHISOYA  & Industrial     \\
			22 & IDBI	      & Financial             &  89 & TNPL	     & Basic Materials       &  156 & BHARATFORG & Industrial     \\
			23 & HDFCBANK   & Financial             &  90 & ITC	       & Consumer Goods        &  157 & GESHIPPING & Industrial     \\
			24 & SBIN       & Financial             &  91 & VSTIND     & Consumer Goods        &  158 & SUNDRMFAST & Industrial     \\
			25 & ORIENTBANK & Financial             &  92 & GODFRYPHLP & Consumer Goods        &  159 & SHYAMTELE  & Telecom        \\
			26 & KARURVYSYA & Financial             &  93 & TATATEA    & Consumer Goods        &  160 & ITI	      & Telecom        \\
			27 & LAKSHVILAS & Financial             &  94 & HARRMALAYA & Consumer Goods        &  161 & HIMACHLFUT & Telecom        \\
			28 & IFCI       & Financial             &  95 & BALRAMCHIN & Consumer Goods        &  162 & MTNL	      & Telecom        \\
			29 & BANKRAJAS  & Financial             &  96 & RAJSREESUG & Consumer Goods        &  163 & BIRLAERIC  & Telecom        \\
			30 & RELCAPITAL & Financial             &  97 & KAKATCEM   & Consumer Goods        &  164 & INDHOTEL   & Services       \\
			31 & CHOLAINV   & Financial             &  98 & SAKHTISUG  & Consumer Goods        &  165 & EIHOTEL    & Services       \\
			32 & FIRSTLEASE & Financial             &  99 & DHAMPURSUG & Consumer Goods        &  166 & ASIANHOTEL & Services       \\
			33 & BAJAUTOFIN & Financial             & 100 & BRITANNIA  & Consumer Goods        &  167 & HOTELEELA  & Services       \\
			34 & SUNDARMFIN & Financial             & 101 & SATNAMOVER & Consumer Goods        &  168 & FLEX	      & Services       \\
			35 & HDFC       & Financial             & 102 & INDSHAVING & Consumer Goods        &  169 & ESSELPACK  & Services       \\
			36 & LICHSGFIN  & Financial             & 103 & MIRCELECTR & Consumer Discretonary &  170 & MAX	      & Services       \\
			37 & CANFINHOME & Financial             & 104 & SURAJDIAMN & Consumer Discretonary &  171 & COSMOFILMS & Services       \\
			38 & GICHSGFIN  & Financial             & 105 & SAMTEL     & Consumer Discretonary &  172 & DABUR	    & Health Care    \\
			39 & TFCILTD    & Financial             & 106 & VDOCONAPPL & Consumer Discretonary &  173 & COLGATE    & Health Care    \\
			40 & TATAELXSI  & Technology            & 107 & VDOCONINTL & Consumer Discretonary &  174 & GLAXO	    & Health Care    \\
			41 & MOSERBAER  & Technology            & 108 & INGERRAND  & Consumer Discretonary &  175 & DRREDDY    & Health Care    \\
			42 & SATYAMCOMP & Technology            & 109 & ELGIEQUIP  & Consumer Discretonary &  176 & CIPLA	    & Health Care    \\
			43 & ROLTA      & Technology            & 110 & KSBPUMPS   & Consumer Discretonary &  177 & RANBAXY    & Health Care    \\
			44 & INFOSYSTCH & Technology            & 111 & NIRMA	     & Consumer Discretonary &  178 & SUNPHARMA  & Health Care    \\
			45 & MASTEK     & Technology            & 112 & VOLTAS     & Consumer Discretonary &  179 & IPCALAB    & Health Care    \\
			46 & WIPRO      & Technology            & 113 & KECINTL    & Consumer Discretonary &  180 & PFIZER     & Health Care    \\
			47 & BEML       & Technology            & 114 & TUBEINVEST & Consumer Discretonary &  181 & EMERCK     & Health Care    \\
			48 & ALFALAVAL  & Technology            & 115 & TITAN	     & Consumer Discretonary &  182 & NICOLASPIR & Health Care    \\
			49 & RIIL       & Technology            & 116 & ABB	       & Industrial            &  183 & SHASUNCHEM & Health Care    \\
			50 & GIPCL      & Energy                & 117 & BHEL	     & Industrial            &  184 & AUROPHARMA & Health Care    \\
			51 & CESC       & Energy                & 118 & THERMAX    & Industrial            &  185 & NATCOPHARM & Health Care    \\
			52 & TATAPOWER  & Energy                & 119 & SIEMENS    & Industrial            &  186 & HINDLEVER  & Miscellaneous  \\
			53 & GUJRATGAS  & Energy                & 120 & CROMPGREAV & Industrial            &  187 & CENTURYTEX & Miscellaneous  \\
			54 & GUJFLUORO  & Energy                & 121 & HEG	       & Industrial            &  188 & EIDPARRY   & Miscellaneous  \\
			55 & HINDOILEXP & Energy                & 122 & ESABINDIA  & Industrial            &  189 & KESORAMIND & Miscellaneous  \\
			56 & ONGC	      & Energy                & 123 & BATAINDIA  & Industrial            &  190 & ADANIEXPO  & Miscellaneous  \\
			57 & COCHINREFN & Energy                & 124 & ASIANPAINT & Industrial            &  191 & ZEETELE    & Miscellaneous  \\
			58 & IPCL	      & Energy                & 125 & ICI	       & Industrial            &  192 & FINCABLES  & Miscellaneous  \\
			59 & FINPIPE    & Energy                & 126 & BERGEPAINT & Industrial            &  193 & RAMANEWSPR & Miscellaneous  \\
			60 & TNPETRO    & Energy                & 127 & GNFC	     & Industrial            &  194 & APOLLOHOSP & Miscellaneous  \\
			61 & SUPPETRO   & Energy                & 128 & NAGARFERT  & Industrial            &  195 & THOMASCOOK & Miscellaneous  \\
			62 & DCW	      & Energy                & 129 & DEEPAKFERT & Industrial            &  196 & POLYPLEX   & Miscellaneous  \\
			63 & CHEMPLAST  & Energy                & 130 & GSFC	     & Industrial            &  197 & BLUEDART   & Miscellaneous  \\
			64 & RELIANCE   & Energy                & 131 & ZUARIAGRO  & Industrial            &  198 & GTCIND     & Miscellaneous  \\
			65 & HINDPETRO  & Energy                & 132 & GODAVRFERT & Industrial            &  199 & TATAVASHIS & Miscellaneous  \\
			66 & BONGAIREFN & Energy                & 133 & ARVINDMILL & Industrial            &  200 & CRISIL     & Miscellaneous  \\
			67 & BPCL	      & Energy                & 134 & RAYMOND    & Industrial            &  201 & INDRAYON   & Miscellaneous  \\
		\end{tabular}                                                                       
	\end{ruledtabular}
\end{table*}
\endgroup
\end{document}